# Demonstration of 17λ × 10 Gb/s C-Band Classical / DV-QKD Co-Existence Over Hollow-Core Fiber Link

F. Honz[(1)], F. Prawits[(1)], O. Alia[(2)], H. Sakr[(3)], T. Bradley[(3)], C. Zhang[(3)], R. Slavík[(3)], F. Poletti[(3)], G. Kanellos[(2)], R. Nejabati[(2)], P. Walther[(4)], D. Simeonidou[(2)], H. Hübel[(1)] and B. Schrenk[(1)]

[(1)] AIT Austrian Institute of Technology, 1210 Vienna, Austria. florian.honz@ait.ac.at
[(2)] University of Bristol, Bristol, United Kingdom.
[(3)] Optoelectronics Research Centre, University of Southampton, Southampton, United Kingdom.
[(4)] University of Vienna, Faculty of Physics, 1090 Vienna, Austria.

**Abstract** *We successfully integrate coherent one-way QKD at 1538 nm in a 7.7 km long hollow-core fiber link with 17 EDFA-boosted C-band data channels from 1540.56 to 1558.17 nm, aggregating a power of 11 dBm. QKD operation proves successful despite the wideband layout of classical channels.*
©2022 The Author(s)

**Introduction**

The security threat of quantum computing to our current communication protocols can be provably countered by employing quantum key distribution (QKD). Significant leaps from lab-scale experiments to field-installable products have resulted in the burgeoning of off-the-shelf QKD solutions, mainly dedicated to securing point-to-point links. In view of co-existence with carrier-grade classical channels that feature a ~90 dB higher per-channel launch than QKD, the practical network integration remains a challenge. Most QKD demonstrations have therefore relied on dark fiber for either the distillation channel or the classical network load. However, this is not a viable option under stringent operational expenditures or in fiber-scarce environments. Recent works [1-5] (Fig. 1a) have therefore investigated the robustness of QKD to impairments associated to the co-propagation of classical channels and QKD (including its classical channel for key distillation). Especially in-band crosstalk noise due to Raman scattering has been identified as a primary impediment that is further propelled by the high dynamic range between classical and quantum power levels [6]. This noise regime would advocate continuous-variable (CV) QKD (●) by virtue of its noise filtering inherent to coherent reception [1-3]. However, CV-QKD cannot maintain secure-key generation over high optical budgets or an extended reach. Alternatively, the remote O-band (□) [4-5] can be dedicated to loss-robust discrete-variable (DV) QKD [7]. However, robustness to classical-channel noise is gained at the expense of reach limitations due to the higher O-band fiber loss. Moreover, the hybrid O/C-band layout does not bode well with emerging ultra-wideband schemes [8].

In this work, we experimentally demonstrate wideband (C-band) classical / 1538-nm DV-QKD co-existence over a metro-scale hollow-core fiber link with greatly reduced Raman scattering. We perform secure-key generation at 65 bit/s while loading the link with 17 data channels aggregating a power of 11 dBm. This corresponds to an 8-dB improvement in co-existing classical power while at the same time increasing the number of co-propagating channels by a factor of 1.7 (Fig. 1a) and making the QKD integration widely independent of the spectral layout of the classical data channels.

**WDM-Robust Integration of Quantum Signal**

When co-propagating classical and quantum channels over the same glass-core fiber, in-band Raman noise induced by the strong classical channels is inevitable. Figure 1b evidences the broadband nature of spontaneous Raman scattering (SpRS) for a standard single-mode fiber (SMF), conducted with a classical and single-photon ($h\nu$) OSA. Measurements for a 1550-nm pump channel show that its SpRS tails reach till the O-band, where the noise contribution eventually falls below $10^5$ photons/s/100pm ($\Lambda$), a value that makes it compatible with QKD [4-5]. To battle this impairment in the C-band at the immediate vicinity of the pump, this work builds on initial findings [9] that employ a hollow-core fiber (HCF) with an air-filled core to mitigate Raman scattering and four-wave mixing. We used a 7.7 km long HCF sample with SMF28 pigtails, described in detail in [10]. As reported in Fig. 1b, the SpRS of the HCF decreases by ~35 dB with respect to the SMF. This permits us to explore a wide region of low Raman noise in the C-band, a band that had been predominantly used with a limited number of artificially attenuated classical channels located very close to the DV quantum channel [7] (■ in Fig. 1a) to ensure a small spectral detuning with a weak Raman noise contribution (Δ in Fig. 1b).

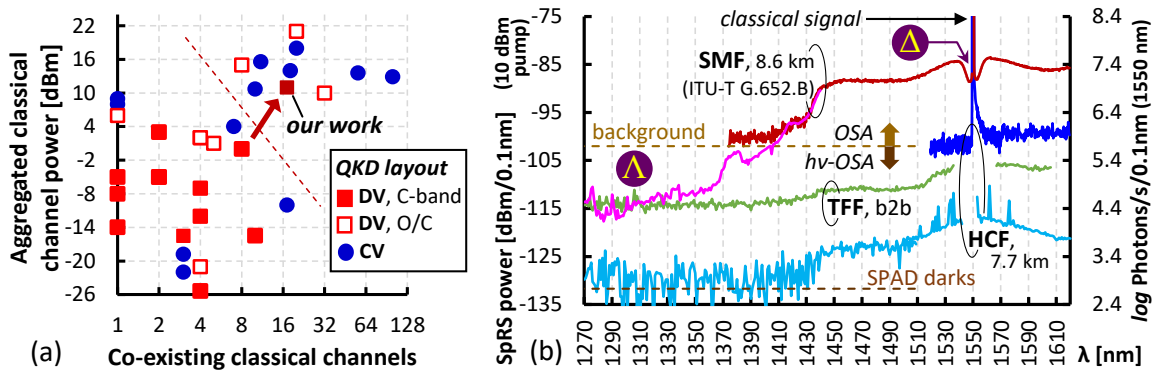

**Fig. 1:** (a) QKD co-existence demonstrations. (b) Spontaneous Raman scattering spectra of SMF and HCF.

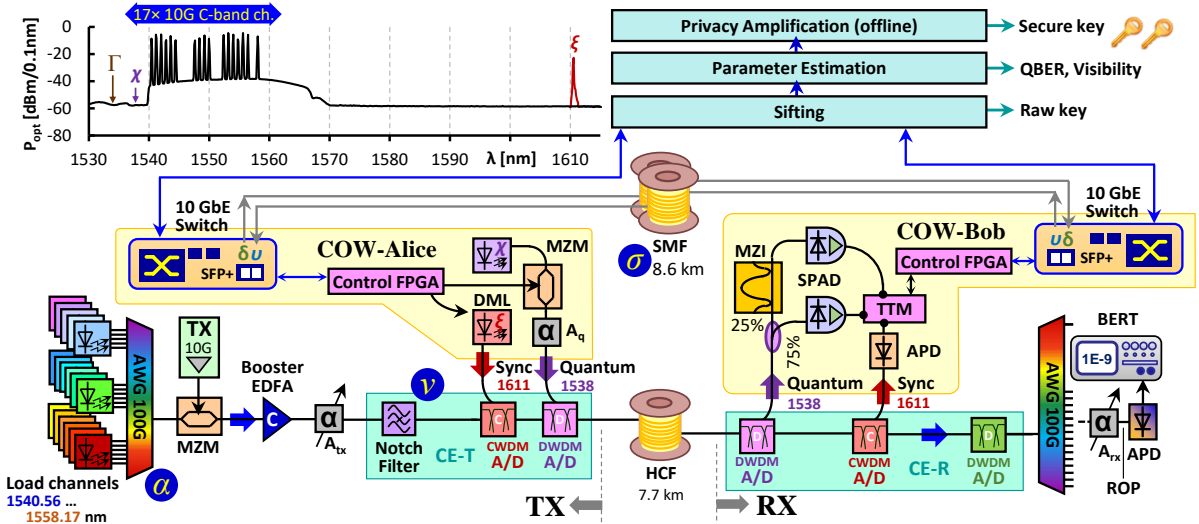

**Fig. 2:** Experimental setup for joint QKD and C-band transmission. The spectral inset presents the transmitted spectrum.

**C-Band DV-QKD / Classical Co-Existence**

Figure 2 presents the experimental setup. The coherent one-way (COW) QKD protocol [11] is implemented at 1538 nm ($\chi$) with a pulse rate of 1 GHz and an average photon number of 0.1 photons/pulse. Framing at 1 MHz is accompanied by a sync pulse at 1611 nm ($\xi$) acting as the local time reference. The QKD receiver employs two free-running InGaAs SPADs (10% efficiency, dark count rate of 620 Hz) to determine either the time of arrival (75% branch) or the phase between consecutive pulses (25% branch). QKD optics are orchestrated through FPGA-based controllers, which perform pattern generation, synchronization and registration of detection events to the key distillation stack. The latter conducts (*i*) sifting of the raw key, (*ii*) parameter estimation yielding QBER and visibility, and (*iii*) offline privacy amplification, eventually resulting in a secure key. For this purpose, a bidirectional 10GbE Ethernet connection is implemented through SFP+ optics in down- ($\delta$, 1554.94 nm) and uplink ($\upsilon$, 1552.52 nm) directions. This bidirectional distillation channel was implemented on a separate SMF pair ($\sigma$). The link load adjacent to the QKD channel consists of 17 intensity-modulated channels in the C-band (see inset in Fig. 2), carrying 10 Gb/s PRBS data and being boosted by a 20-dBm C-band EDFA. Co-existence with QKD necessitates spectral conditioning for the classical signal spectrum. This includes (*i*) filtering of the amplified spontaneous emission (ASE) tails of the laser lines, accomplished through the arrayed waveguide grating (AWG, $\alpha$) used to multiplex the optical sources and (*ii*) suppression of the ASE of the EDFA through a notch filter with a rejection of >95 dB (Fig. 3a, $\nu$), cleaning out the quantum channel at 1538 nm. The launch power of the classical load channels has been controlled (attenuator $A_{tx}$) to investigate the co-existence limits. The quantum signal and the corresponding frame sync at 1611 nm are lastly multiplexed to the classical signal compound. This ensures minimal multiplexing losses of 0.76 dB for the fragile quantum signal (Fig. 3a, $\chi$), while the notching and express feed-through of the classical channels at the transmitter-side co-existence element (CE-T) amount to 6.3 dB ($\iota$). It shall be stressed that the notch at the quantum channel $\chi$ is not visible in Fig. 2 due to additional scattering arising at the OSA grating ($\Gamma$). At the receiving end of the link, a co-existence element (CE-R) drops the quantum and QKD sync channels before the classical channels (Fig. 3a). The downlink has been demultiplexed and received by an APD receiver in order to prove the quality of signal reception through BER measurements ($A_{rx}$) as a function of the received optical power (ROP).

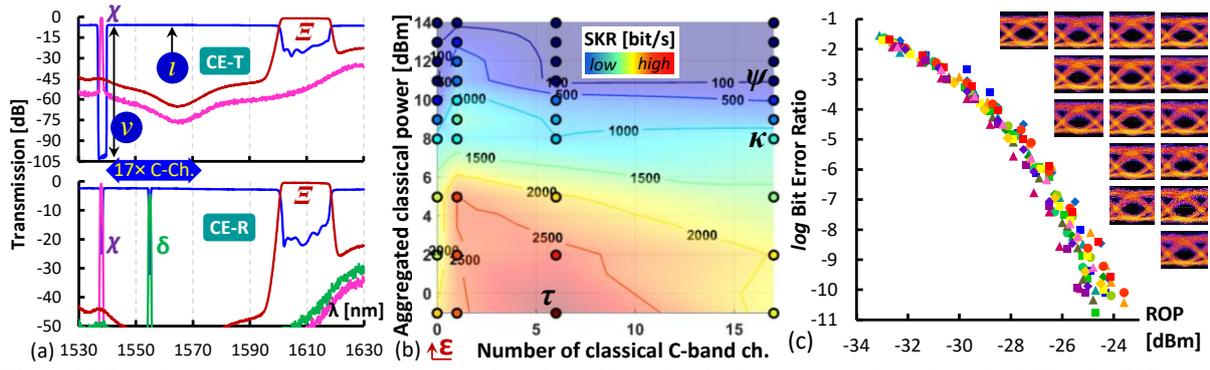

**Fig. 3:** (a) Co-existence elements at transmitter and receiver with notch $\nu$ for the quantum channel $\chi$ and add/drop for QKD sync $\Xi$. (b) Secure-key rate under presence of classical channels. (c) BER and eye diagrams for the data channels.

**Results: Secure-Key Rate and BER**
First, the secure-key rate (SKR) has been evaluated as a function of the aggregated classical power and channel count at the C-band (Fig. 3b). The SKR performance is primarily determined by the aggregated classical power due to the flat and low Raman scattering profile of the HCF (Fig. 1b), making the SKR widely independent of the spectral layout. A SKR of 65 bit/s can still be obtained for 17 co-propagating channels aggregating 11 dBm ($\psi$). This proves the robustness of DV-QKD in combination with the HCF to a WDM feed of data channels spanning over 17.6 nm of the C-band fiber spectrum, without being restricted to a narrow (few nm) spectral slice close to the quantum channel as it applies to a SMF ($\Delta$ in Fig. 1b). We noticed a low background noise even for a back-to-back configuration without HCF, which was eventually identified to originate from scattering at SMF-pigtailed thin-film WDM filters employed at the CE-T element (TFF in Fig. 1b). This imposes an upper bound for the compatible launch power. The SKR expands to 1.2 kb/s ($\kappa$) for a weaker launch power of 8 dBm, leading to a QBER of 1.2%. If we apply the NIST recommendation for AES key renewal, meaning that a single 256-bit long AES key should only be used for encrypting a maximum of 64 Gbyte of data, these SKRs can already accommodate data security for the total link capacity of 17×10 Gb/s. A weaker classical signal launch of -1 dBm (using only the closest six C-band channels) permits a higher SKR of up to 3.1 kb/s ($\tau$), owing to the low intrinsic QBER of 0.67% of the COW-QKD system. The drop of the SKR to 1.5 – 2 kb/s without classical channels ($\varepsilon$ in Fig. 3b) is attributed to the considerably elevated ASE background under no-load conditions for the booster EDFA, which contaminates the quantum channel. Finally, we acquired the BER performance for the classical channels (Fig. 3c), showing a sensitivity better than -23.6 dBm at a BER of $10^{-10}$ and a spread of 1.8 dB between best and worst channel. All eye diagrams were clearly open.

**Conclusion**
We have demonstrated secure-key generation at a 1538-nm quantum channel, in co-propagation with a C-band DWDM feed from 1540.56 to 1558.17 nm. We obtained a SKR of 65 bit/s for an aggregated classical power of 11 dBm, proving that a secure key can be established despite a spectrally widespread C-band DWDM traffic standing in co-existence adjacent to the C-band QKD channel. This accomplishment corresponds to a massive 19 dB increase in terms of classical power × optical bandwidth product compared to earlier DV-QKD works and enables the QKD integration in high-capacity, long-range metro-core links. The mitigation of Raman scattering in SMF spans, as accomplished using a novel HCF, now points to the need for upgrading fiber-optic components as well: We found thin-film add/drop multiplexers combining quantum and 90 dB stronger classical signals to prevent further scaling in terms of higher SKR and stronger classical launch. This would require WDM components with hollow-core layout to become available. Together with sub-DWDM filtering at the quantum channel, we expect classical levels beyond 20 dBm to be compatible.

**Acknowledgements**

This work received funding from the EU Horizon-2020 programme (grant 820474) and the Austrian FFG (grant 881112).